\pdfoutput=1
\documentclass[11pt,twoside,a4paper,cmspaper,final,collab]{cms-tdr}
\def\svnVersion{18744:18760}\def\cmsCernNo{2010-035}\def\cmsCernDate{2010/10/04}\def\cmsMessage{Submitted to Physical Review Letters}

\begin{document}\cmsNoteHeader{EXO-10-010}
%
%
%

%
%
\hyphenation{env-iron-men-tal}
\hyphenation{had-ron-i-za-tion}
\hyphenation{cal-or-i-me-ter}
\hyphenation{de-vices}
%
\RCS$Revision: 18760 $
\RCS$HeadURL: svn+ssh://alverson@svn.cern.ch/reps/tdr2/papers/EXO-10-010/trunk/EXO-10-010.tex $
\RCS$Id: EXO-10-010.tex 18760 2010-10-19 05:54:01Z alverson $
%
%
%

\providecommand {\etal}{\mbox{et al.}\xspace} 
\providecommand {\ie}{\mbox{i.e.}\xspace}     
\providecommand {\eg}{\mbox{e.g.}\xspace}     
\providecommand {\etc}{\mbox{etc.}\xspace}     
\providecommand {\vs}{\mbox{\sl vs.}\xspace}      
\providecommand {\mdash}{\ensuremath{\mathrm{-}}} 

\providecommand {\Lone}{Level-1\xspace} 
\providecommand {\Ltwo}{Level-2\xspace}
\providecommand {\Lthree}{Level-3\xspace}

\providecommand{\ACERMC} {\textsc{AcerMC}\xspace}
\providecommand{\ALPGEN} {{\textsc{alpgen}}\xspace}
\providecommand{\CHARYBDIS} {{\textsc{charybdis}}\xspace}
\providecommand{\CMKIN} {\textsc{cmkin}\xspace}
\providecommand{\CMSIM} {{\textsc{cmsim}}\xspace}
\providecommand{\CMSSW} {{\textsc{cmssw}}\xspace}
\providecommand{\COBRA} {{\textsc{cobra}}\xspace}
\providecommand{\COCOA} {{\textsc{cocoa}}\xspace}
\providecommand{\COMPHEP} {\textsc{CompHEP}\xspace}
\providecommand{\EVTGEN} {{\textsc{evtgen}}\xspace}
\providecommand{\FAMOS} {{\textsc{famos}}\xspace}
\providecommand{\GARCON} {\textsc{garcon}\xspace}
\providecommand{\GARFIELD} {{\textsc{garfield}}\xspace}
\providecommand{\GEANE} {{\textsc{geane}}\xspace}
\providecommand{\GEANTfour} {{\textsc{geant4}}\xspace}
\providecommand{\GEANTthree} {{\textsc{geant3}}\xspace}
\providecommand{\GEANT} {{\textsc{geant}}\xspace}
\providecommand{\HDECAY} {\textsc{hdecay}\xspace}
\providecommand{\HERWIG} {{\textsc{herwig}}\xspace}
\providecommand{\HIGLU} {{\textsc{higlu}}\xspace}
\providecommand{\HIJING} {{\textsc{hijing}}\xspace}
\providecommand{\IGUANA} {\textsc{iguana}\xspace}
\providecommand{\ISAJET} {{\textsc{isajet}}\xspace}
\providecommand{\ISAPYTHIA} {{\textsc{isapythia}}\xspace}
\providecommand{\ISASUGRA} {{\textsc{isasugra}}\xspace}
\providecommand{\ISASUSY} {{\textsc{isasusy}}\xspace}
\providecommand{\ISAWIG} {{\textsc{isawig}}\xspace}
\providecommand{\MADGRAPH} {\textsc{MadGraph}\xspace}
\providecommand{\MCATNLO} {\textsc{mc@nlo}\xspace}
\providecommand{\MCFM} {\textsc{mcfm}\xspace}
\providecommand{\MILLEPEDE} {{\textsc{millepede}}\xspace}
\providecommand{\ORCA} {{\textsc{orca}}\xspace}
\providecommand{\OSCAR} {{\textsc{oscar}}\xspace}
\providecommand{\PHOTOS} {\textsc{photos}\xspace}
\providecommand{\PROSPINO} {\textsc{prospino}\xspace}
\providecommand{\PYTHIA} {{\textsc{pythia}}\xspace}
\providecommand{\SHERPA} {{\textsc{sherpa}}\xspace}
\providecommand{\TAUOLA} {\textsc{tauola}\xspace}
\providecommand{\TOPREX} {\textsc{TopReX}\xspace}
\providecommand{\XDAQ} {{\textsc{xdaq}}\xspace}

\providecommand {\DZERO}{D\O\xspace}     


\providecommand{\de}{\ensuremath{^\circ}}
\providecommand{\ten}[1]{\ensuremath{\times \text{10}^\text{#1}}}
\providecommand{\unit}[1]{\ensuremath{\text{\,#1}}\xspace}
\providecommand{\mum}{\ensuremath{\,\mu\text{m}}\xspace}
\providecommand{\micron}{\ensuremath{\,\mu\text{m}}\xspace}
\providecommand{\cm}{\ensuremath{\,\text{cm}}\xspace}
\providecommand{\mm}{\ensuremath{\,\text{mm}}\xspace}
\providecommand{\mus}{\ensuremath{\,\mu\text{s}}\xspace}
\providecommand{\keV}{\ensuremath{\,\text{ke\hspace{-.08em}V}}\xspace}
\providecommand{\MeV}{\ensuremath{\,\text{Me\hspace{-.08em}V}}\xspace}
\providecommand{\GeV}{\ensuremath{\,\text{Ge\hspace{-.08em}V}}\xspace}
\providecommand{\TeV}{\ensuremath{\,\text{Te\hspace{-.08em}V}}\xspace}
\providecommand{\PeV}{\ensuremath{\,\text{Pe\hspace{-.08em}V}}\xspace}
\providecommand{\keVc}{\ensuremath{{\,\text{ke\hspace{-.08em}V\hspace{-0.16em}/\hspace{-0.08em}}c}}\xspace}
\providecommand{\MeVc}{\ensuremath{{\,\text{Me\hspace{-.08em}V\hspace{-0.16em}/\hspace{-0.08em}}c}}\xspace}
\providecommand{\GeVc}{\ensuremath{{\,\text{Ge\hspace{-.08em}V\hspace{-0.16em}/\hspace{-0.08em}}c}}\xspace}
\providecommand{\TeVc}{\ensuremath{{\,\text{Te\hspace{-.08em}V\hspace{-0.16em}/\hspace{-0.08em}}c}}\xspace}
\providecommand{\keVcc}{\ensuremath{{\,\text{ke\hspace{-.08em}V\hspace{-0.16em}/\hspace{-0.08em}}c^\text{2}}}\xspace}
\providecommand{\MeVcc}{\ensuremath{{\,\text{Me\hspace{-.08em}V\hspace{-0.16em}/\hspace{-0.08em}}c^\text{2}}}\xspace}
\providecommand{\GeVcc}{\ensuremath{{\,\text{Ge\hspace{-.08em}V\hspace{-0.16em}/\hspace{-0.08em}}c^\text{2}}}\xspace}
\providecommand{\TeVcc}{\ensuremath{{\,\text{Te\hspace{-.08em}V\hspace{-0.16em}/\hspace{-0.08em}}c^\text{2}}}\xspace}

\providecommand{\pbinv} {\mbox{\ensuremath{\,\text{pb}^\text{$-$1}}}\xspace}
\providecommand{\fbinv} {\mbox{\ensuremath{\,\text{fb}^\text{$-$1}}}\xspace}
\providecommand{\nbinv} {\mbox{\ensuremath{\,\text{nb}^\text{$-$1}}}\xspace}
\providecommand{\percms}{\ensuremath{\,\text{cm}^\text{$-$2}\,\text{s}^\text{$-$1}}\xspace}
\providecommand{\lumi}{\ensuremath{\mathcal{L}}\xspace}
\providecommand{\Lumi}{\ensuremath{\mathcal{L}}\xspace}
%
\providecommand{\LvLow}  {\ensuremath{\mathcal{L}=\text{10}^\text{32}\,\text{cm}^\text{$-$2}\,\text{s}^\text{$-$1}}\xspace}
\providecommand{\LLow}   {\ensuremath{\mathcal{L}=\text{10}^\text{33}\,\text{cm}^\text{$-$2}\,\text{s}^\text{$-$1}}\xspace}
\providecommand{\lowlumi}{\ensuremath{\mathcal{L}=\text{2}\times \text{10}^\text{33}\,\text{cm}^\text{$-$2}\,\text{s}^\text{$-$1}}\xspace}
\providecommand{\LMed}   {\ensuremath{\mathcal{L}=\text{2}\times \text{10}^\text{33}\,\text{cm}^\text{$-$2}\,\text{s}^\text{$-$1}}\xspace}
\providecommand{\LHigh}  {\ensuremath{\mathcal{L}=\text{10}^\text{34}\,\text{cm}^\text{$-$2}\,\text{s}^\text{$-$1}}\xspace}
\providecommand{\hilumi} {\ensuremath{\mathcal{L}=\text{10}^\text{34}\,\text{cm}^\text{$-$2}\,\text{s}^\text{$-$1}}\xspace}


\providecommand{\zp}{\ensuremath{\mathrm{Z}^\prime}\xspace}


\providecommand{\kt}{\ensuremath{k_{\mathrm{T}}}\xspace}
\providecommand{\BC}{\ensuremath{\mathrm{B_{c}}}\xspace}
\providecommand{\bbarc}{\ensuremath{\mathrm{\overline{b}c}}\xspace}
\providecommand{\bbbar}{\ensuremath{\mathrm{b\overline{b}}}\xspace}
\providecommand{\ccbar}{\ensuremath{\mathrm{c\overline{c}}}\xspace}
\providecommand{\JPsi}{\ensuremath{\mathrm{J}\hspace{-.08em}/\hspace{-.14em}\psi}\xspace}
\providecommand{\bspsiphi}{\ensuremath{\mathrm{B_s} \to \JPsi\, \phi}\xspace}
\providecommand{\AFB}{\ensuremath{A_\text{FB}}\xspace}
\providecommand{\EE}{\ensuremath{\mathrm{e^+e^-}}\xspace}
\providecommand{\MM}{\ensuremath{\mu^+\mu^-}\xspace}
\providecommand{\TT}{\ensuremath{\tau^+\tau^-}\xspace}
\providecommand{\wangle}{\ensuremath{\sin^{2}\theta_{\text{eff}}^\text{lept}(M^2_\mathrm{Z})}\xspace}
\providecommand{\ttbar}{\ensuremath{\mathrm{t\overline{t}}}\xspace}
\providecommand{\stat}{\ensuremath{\,\text{(stat.)}}\xspace}
\providecommand{\syst}{\ensuremath{\,\text{(syst.)}}\xspace}

\providecommand{\HGG}{\ensuremath{\mathrm{H}\to\gamma\gamma}}
\providecommand{\gev}{\GeV}
\providecommand{\GAMJET}{\ensuremath{\gamma + \text{jet}}}
\providecommand{\PPTOJETS}{\ensuremath{\mathrm{pp}\to\text{jets}}}
\providecommand{\PPTOGG}{\ensuremath{\mathrm{pp}\to\gamma\gamma}}
\providecommand{\PPTOGAMJET}{\ensuremath{\mathrm{pp}\to\gamma + \mathrm{jet}}}
\providecommand{\MH}{\ensuremath{M_{\mathrm{H}}}}
\providecommand{\RNINE}{\ensuremath{R_\mathrm{9}}}
\providecommand{\DR}{\ensuremath{\Delta R}}


\providecommand{\PT}{\ensuremath{p_{\mathrm{T}}}\xspace}
\providecommand{\pt}{\ensuremath{p_{\mathrm{T}}}\xspace}
\providecommand{\ET}{\ensuremath{E_{\mathrm{T}}}\xspace}
\providecommand{\HT}{\ensuremath{H_{\mathrm{T}}}\xspace}
\providecommand{\et}{\ensuremath{E_{\mathrm{T}}}\xspace}
\providecommand{\Em}{\ensuremath{E\hspace{-0.6em}/}\xspace}
\providecommand{\Pm}{\ensuremath{p\hspace{-0.5em}/}\xspace}
\providecommand{\PTm}{\ensuremath{{p}_\mathrm{T}\hspace{-1.02em}/}\xspace}
\providecommand{\PTslash}{\ensuremath{{p}_\mathrm{T}\hspace{-1.02em}/}\xspace}
\providecommand{\ETm}{\ensuremath{E_{\mathrm{T}}^{\text{miss}}}\xspace}
\providecommand{\ETslash}{\ensuremath{E_{\mathrm{T}}\hspace{-1.1em}/}\xspace}
\providecommand{\MET}{\ensuremath{E_{\mathrm{T}}^{\text{miss}}}\xspace}
\providecommand{\ETmiss}{\ensuremath{E_{\mathrm{T}}^{\text{miss}}}\xspace}
\providecommand{\VEtmiss}{\ensuremath{{\vec E}_{\mathrm{T}}^{\text{miss}}}\xspace}

\providecommand{\dd}[2]{\ensuremath{\frac{\mathrm{d} #1}{\mathrm{d} #2}}}

%

\providecommand{\ga}{\ensuremath{\gtrsim}}
\providecommand{\la}{\ensuremath{\lesssim}}
\providecommand{\swsq}{\ensuremath{\sin^2\theta_\mathrm{W}}\xspace}
\providecommand{\cwsq}{\ensuremath{\cos^2\theta_\mathrm{W}}\xspace}
\providecommand{\tanb}{\ensuremath{\tan\beta}\xspace}
\providecommand{\tanbsq}{\ensuremath{\tan^{2}\beta}\xspace}
\providecommand{\sidb}{\ensuremath{\sin 2\beta}\xspace}
\providecommand{\alpS}{\ensuremath{\alpha_S}\xspace}
\providecommand{\alpt}{\ensuremath{\tilde{\alpha}}\xspace}

\providecommand{\QL}{\ensuremath{\mathrm{Q}_\mathrm{L}}\xspace}
\providecommand{\sQ}{\ensuremath{\tilde{\mathrm{Q}}}\xspace}
\providecommand{\sQL}{\ensuremath{\tilde{\mathrm{Q}}_\mathrm{L}}\xspace}
\providecommand{\ULC}{\ensuremath{\mathrm{U}_\mathrm{L}^\mathrm{C}}\xspace}
\providecommand{\sUC}{\ensuremath{\tilde{\mathrm{U}}^\mathrm{C}}\xspace}
\providecommand{\sULC}{\ensuremath{\tilde{\mathrm{U}}_\mathrm{L}^\mathrm{C}}\xspace}
\providecommand{\DLC}{\ensuremath{\mathrm{D}_\mathrm{L}^\mathrm{C}}\xspace}
\providecommand{\sDC}{\ensuremath{\tilde{\mathrm{D}}^\mathrm{C}}\xspace}
\providecommand{\sDLC}{\ensuremath{\tilde{\mathrm{D}}_\mathrm{L}^\mathrm{C}}\xspace}
\providecommand{\LL}{\ensuremath{\mathrm{L}_\mathrm{L}}\xspace}
\providecommand{\sL}{\ensuremath{\tilde{\mathrm{L}}}\xspace}
\providecommand{\sLL}{\ensuremath{\tilde{\mathrm{L}}_\mathrm{L}}\xspace}
\providecommand{\ELC}{\ensuremath{\mathrm{E}_\mathrm{L}^\mathrm{C}}\xspace}
\providecommand{\sEC}{\ensuremath{\tilde{\mathrm{E}}^\mathrm{C}}\xspace}
\providecommand{\sELC}{\ensuremath{\tilde{\mathrm{E}}_\mathrm{L}^\mathrm{C}}\xspace}
\providecommand{\sEL}{\ensuremath{\tilde{\mathrm{E}}_\mathrm{L}}\xspace}
\providecommand{\sER}{\ensuremath{\tilde{\mathrm{E}}_\mathrm{R}}\xspace}
\providecommand{\sFer}{\ensuremath{\tilde{\mathrm{f}}}\xspace}
\providecommand{\sQua}{\ensuremath{\tilde{\mathrm{q}}}\xspace}
\providecommand{\sUp}{\ensuremath{\tilde{\mathrm{u}}}\xspace}
\providecommand{\suL}{\ensuremath{\tilde{\mathrm{u}}_\mathrm{L}}\xspace}
\providecommand{\suR}{\ensuremath{\tilde{\mathrm{u}}_\mathrm{R}}\xspace}
\providecommand{\sDw}{\ensuremath{\tilde{\mathrm{d}}}\xspace}
\providecommand{\sdL}{\ensuremath{\tilde{\mathrm{d}}_\mathrm{L}}\xspace}
\providecommand{\sdR}{\ensuremath{\tilde{\mathrm{d}}_\mathrm{R}}\xspace}
\providecommand{\sTop}{\ensuremath{\tilde{\mathrm{t}}}\xspace}
\providecommand{\stL}{\ensuremath{\tilde{\mathrm{t}}_\mathrm{L}}\xspace}
\providecommand{\stR}{\ensuremath{\tilde{\mathrm{t}}_\mathrm{R}}\xspace}
\providecommand{\stone}{\ensuremath{\tilde{\mathrm{t}}_1}\xspace}
\providecommand{\sttwo}{\ensuremath{\tilde{\mathrm{t}}_2}\xspace}
\providecommand{\sBot}{\ensuremath{\tilde{\mathrm{b}}}\xspace}
\providecommand{\sbL}{\ensuremath{\tilde{\mathrm{b}}_\mathrm{L}}\xspace}
\providecommand{\sbR}{\ensuremath{\tilde{\mathrm{b}}_\mathrm{R}}\xspace}
\providecommand{\sbone}{\ensuremath{\tilde{\mathrm{b}}_1}\xspace}
\providecommand{\sbtwo}{\ensuremath{\tilde{\mathrm{b}}_2}\xspace}
\providecommand{\sLep}{\ensuremath{\tilde{\mathrm{l}}}\xspace}
\providecommand{\sLepC}{\ensuremath{\tilde{\mathrm{l}}^\mathrm{C}}\xspace}
\providecommand{\sEl}{\ensuremath{\tilde{\mathrm{e}}}\xspace}
\providecommand{\sElC}{\ensuremath{\tilde{\mathrm{e}}^\mathrm{C}}\xspace}
\providecommand{\seL}{\ensuremath{\tilde{\mathrm{e}}_\mathrm{L}}\xspace}
\providecommand{\seR}{\ensuremath{\tilde{\mathrm{e}}_\mathrm{R}}\xspace}
\providecommand{\snL}{\ensuremath{\tilde{\nu}_L}\xspace}
\providecommand{\sMu}{\ensuremath{\tilde{\mu}}\xspace}
\providecommand{\sNu}{\ensuremath{\tilde{\nu}}\xspace}
\providecommand{\sTau}{\ensuremath{\tilde{\tau}}\xspace}
\providecommand{\Glu}{\ensuremath{\mathrm{g}}\xspace}
\providecommand{\sGlu}{\ensuremath{\tilde{\mathrm{g}}}\xspace}
\providecommand{\Wpm}{\ensuremath{\mathrm{W}^{\pm}}\xspace}
\providecommand{\sWpm}{\ensuremath{\tilde{\mathrm{W}}^{\pm}}\xspace}
\providecommand{\Wz}{\ensuremath{\mathrm{W}^{0}}\xspace}
\providecommand{\sWz}{\ensuremath{\tilde{\mathrm{W}}^{0}}\xspace}
\providecommand{\sWino}{\ensuremath{\tilde{\mathrm{W}}}\xspace}
\providecommand{\Bz}{\ensuremath{\mathrm{B}^{0}}\xspace}
\providecommand{\sBz}{\ensuremath{\tilde{\mathrm{B}}^{0}}\xspace}
\providecommand{\sBino}{\ensuremath{\tilde{\mathrm{B}}}\xspace}
\providecommand{\Zz}{\ensuremath{\mathrm{Z}^{0}}\xspace}
\providecommand{\sZino}{\ensuremath{\tilde{\mathrm{Z}}^{0}}\xspace}
\providecommand{\sGam}{\ensuremath{\tilde{\gamma}}\xspace}
\providecommand{\chiz}{\ensuremath{\tilde{\chi}^{0}}\xspace}
\providecommand{\chip}{\ensuremath{\tilde{\chi}^{+}}\xspace}
\providecommand{\chim}{\ensuremath{\tilde{\chi}^{-}}\xspace}
\providecommand{\chipm}{\ensuremath{\tilde{\chi}^{\pm}}\xspace}
\providecommand{\Hone}{\ensuremath{\mathrm{H}_\mathrm{d}}\xspace}
\providecommand{\sHone}{\ensuremath{\tilde{\mathrm{H}}_\mathrm{d}}\xspace}
\providecommand{\Htwo}{\ensuremath{\mathrm{H}_\mathrm{u}}\xspace}
\providecommand{\sHtwo}{\ensuremath{\tilde{\mathrm{H}}_\mathrm{u}}\xspace}
\providecommand{\sHig}{\ensuremath{\tilde{\mathrm{H}}}\xspace}
\providecommand{\sHa}{\ensuremath{\tilde{\mathrm{H}}_\mathrm{a}}\xspace}
\providecommand{\sHb}{\ensuremath{\tilde{\mathrm{H}}_\mathrm{b}}\xspace}
\providecommand{\sHpm}{\ensuremath{\tilde{\mathrm{H}}^{\pm}}\xspace}
\providecommand{\hz}{\ensuremath{\mathrm{h}^{0}}\xspace}
\providecommand{\Hz}{\ensuremath{\mathrm{H}^{0}}\xspace}
\providecommand{\Az}{\ensuremath{\mathrm{A}^{0}}\xspace}
\providecommand{\Hpm}{\ensuremath{\mathrm{H}^{\pm}}\xspace}
\providecommand{\sGra}{\ensuremath{\tilde{\mathrm{G}}}\xspace}
\providecommand{\mtil}{\ensuremath{\tilde{m}}\xspace}
\providecommand{\rpv}{\ensuremath{\rlap{\kern.2em/}R}\xspace}
\providecommand{\LLE}{\ensuremath{LL\bar{E}}\xspace}
\providecommand{\LQD}{\ensuremath{LQ\bar{D}}\xspace}
\providecommand{\UDD}{\ensuremath{\overline{UDD}}\xspace}
\providecommand{\Lam}{\ensuremath{\lambda}\xspace}
\providecommand{\Lamp}{\ensuremath{\lambda'}\xspace}
\providecommand{\Lampp}{\ensuremath{\lambda''}\xspace}
\providecommand{\spinbd}[2]{\ensuremath{\bar{#1}_{\dot{#2}}}\xspace}

\providecommand{\MD}{\ensuremath{{M_\mathrm{D}}}\xspace}
\providecommand{\Mpl}{\ensuremath{{M_\mathrm{Pl}}}\xspace}
\providecommand{\Rinv} {\ensuremath{{R}^{-1}}\xspace} 
\cmsNoteHeader{EXO-10-010} 
\title{Search for Dijet Resonances in 7 TeV $pp$ Collisions at CMS}

\address[cern]{CERN}
\address[neu]{Northeastern University}
\author[cern]{The CMS Collaboration}

\date{\today}

\abstract{
A search for narrow resonances in the dijet mass spectrum
is performed using data corresponding to an integrated luminosity of 2.9 pb$^{-1}$ collected
by the CMS experiment at the Large Hadron Collider. Upper limits at the 95\%
confidence level are presented on the product of the resonance cross
section, branching fraction into dijets, and acceptance, separately for decays into
quark-quark, quark-gluon, or gluon-gluon pairs. The data
exclude new particles predicted in the following models at the
95\% confidence level: string resonances, with mass less than 2.50~TeV, excited quarks, with mass less than 1.58~TeV,
and axigluons, colorons, and $\mbox{E}_6$ diquarks, in specific mass intervals. This extends previously
published limits on these models.}

\hypersetup{%
pdfauthor={CMS Collaboration},%
pdftitle={Search for Dijet Resonances in 7 TeV $pp$ Collisions at CMS},%
pdfsubject={CMS},%
pdfkeywords={CMS, physics, software, computing}}

\maketitle 


Two or more energetic jets arise in proton-proton collisions
when partons are scattered with large transverse
momenta, \pt.  The invariant mass spectrum of the two jets with
largest \pt (dijets) falls steeply and smoothly, as predicted by quantum
chromodynamics (QCD). Many extensions of the standard model predict the existence of new massive objects
that couple to quarks ($q$) and gluons ($g$), and result in resonant structures
in the
dijet mass.  In this Letter we report a search for narrow
resonances in the dijet mass spectrum, measured with the CMS detector~\cite{refCMS} at the LHC, at a proton-proton
collision energy of $\sqrt{s}=7$~TeV.

In addition to this generic search, we search for
narrow s-channel dijet resonances from eight specific models.
First, string resonances ($S$), which are Regge excitations of quarks and gluons in string
theory, with multiple mass-degenerate spin states and
quantum numbers~\cite{Anchordoqui:2008di,Cullen:2000ef}; string resonances with mass
$\sim 2$ TeV are expected to decay predominantly to $qg$ (91\%) with small amounts of $gg$ (5.5\%) and
$q\bar{q}$ (3.5\%).
Second, mass-degenerate excited quarks ($q^*$), which decay to
$qg$, predicted if quarks are composite~\cite{ref_qstar}; the compositeness scale is set to be equal to the mass of the excited quark.
Third, axial vector particles called axigluons ($A$), which decay to  $q\bar{q}$,
predicted in a model where the symmetry group $\hbox{SU(3)}$ of QCD is replaced by the
chiral symmetry ${\hbox{SU(3)}_{\rm L}} \times {\hbox{SU(3)}_{\rm R}}$~\cite{ref_axi}.
Fourth, color-octet colorons ($C$), also decaying to  $q\bar{q}$, predicted by the flavor-universal
coloron model embedding the $\hbox{SU(3)}$ symmetry of QCD in
a larger gauge group~\cite{ref_coloron}.
Fifth, scalar diquarks ($D$), which decay to $qq$ and $\bar{q}\bar{q}$, predicted by a grand unified theory based on
the $\mbox{E}_6$ gauge~\cite{ref_diquark}.
Sixth, Randall-Sundrum (RS) gravitons ($G$), which decay to $q\bar{q}$ and $gg$, predicted in the RS model
of extra dimensions~\cite{ref_rsg}; the value of the dimensionless coupling
$\kappa/\overline{M}_{\rm Pl}$ is chosen to be 0.1.
Seventh and eighth, new
gauge bosons ($W^\prime$ and $Z^\prime$), which decay to $q\bar{q}$,
predicted by models that propose new gauge symmetries~\cite{ref_gauge};
the $W^\prime$ and $Z^\prime$ resonances
are assumed to have standard-model-like couplings.

A detailed description of the CMS experiment can be found
elsewhere~\cite{refCMS}.
The CMS coordinate system has the origin at the center of the detector.
The $z$-axis points along the direction of the counterclockwise beam, with the transverse
plane perpendicular to the beam; $\phi$ is the azimuthal
angle in radians, $\theta$ is the polar angle and the pseudorapidity is
$\eta \equiv -\ln(\tan[\theta/2])$.
The central feature of the CMS apparatus is a superconducting solenoid of 6 m
internal diameter.
Within the field volume are the silicon pixel and strip tracker ($|\eta|<2.4$), and the barrel and
endcap calorimeters ($|\eta|<3$): a lead tungstate crystal electromagnetic calorimeter (ECAL) and
a brass-scintillator hadronic calorimeter (HCAL).
Outside the field volume, in the forward region, there is
an iron-quartz fiber calorimeter ($3<|\eta|<5$).
The ECAL and HCAL cells are grouped into towers, projecting radially outward from
the origin, for triggering purposes and to facilitate jet reconstruction.
In the region
$|\eta|<1.74$ these projective calorimeter towers have segmentation
$\Delta\eta = \Delta\phi = 0.087$; the $\eta$ and $\phi$ width
increases at higher values of $\eta$.
The energy depositions measured in the ECAL and the
HCAL within each projective tower are summed to find the calorimeter tower energy.

The integrated luminosity of the data sample selected for this
analysis is $2.9 \pm 0.3$~pb$^{-1}$~\cite{EWK-10-004-PAS}.  A single-jet trigger is used in both the online
hardware-level (L1) and the software-level (HLT) of the trigger system~\cite{refCMS} to select an unprescaled
sample of events with a nominal jet transverse energy threshold at the HLT of 50~GeV.
The trigger efficiency for this analysis is measured from the data
to be larger than 99.5\% for dijet masses above 220~GeV.

Jets are reconstructed using the anti-$k_T$ algorithm~\cite{1126-6708-2008-04-063} with a distance
parameter $R=0.7$. The reconstructed jet energy $E$ is defined as the scalar
sum of the calorimeter tower energies inside the jet. The jet momentum $\vec{p}$ is the
corresponding vector sum of the tower energies using the tower directions.
The $E$ and $\vec{p}$ of a reconstructed jet are corrected as a function of \pt and $\eta$ for the
non-linearity and inhomogeneity of the calorimeter response.
The correction is between 43\% and 15\% for jets with corrected \pt between $0.1$ and
$1.0$ TeV in the region $|\eta|<1.3$.
The jet energy corrections were determined and validated using simulations, test beam data,
and collision data~\cite{JME-10-003-PAS}.

The dijet system is composed of the
two jets with the highest \pt in an event (leading jets).
We require that the pseudorapidity separation
of the two leading jets, $\Delta\eta=\eta_1-\eta_2$, satisfies
$|\Delta\eta|<1.3$, and that both jets be in the region
$|\eta|<2.5$. These $\eta$ cuts maximize the search sensitivity for
isotropic decays of dijet resonances in the presence of QCD background.
The dijet mass is given by
$m=\sqrt{(E_1 + E_2)^2 - (\vec{p}_1 + \vec{p}_2)^2}$.
We select events with $m>220$~GeV without any requirements on jet \pt.

To remove possible instrumental and non-collision
backgrounds in the selected sample, the following selections are made.
Events are
required to have a reconstructed primary vertex within $|z|<24$ cm.
For jets, at least 1\% of the jet energy must be
detected in the ECAL, at most 98\% can be measured in a single
photodetection device of the HCAL readout, and at most 90\%
can be measured in a single cell.
These criteria, which are fully efficient for dijets, remove 0.1\% of the events passing the pseudorapidity constraints
and the dijet mass threshold.

\begin{figure}[hbtp]
  \begin{center}
    \includegraphics[width=1.0\columnwidth]{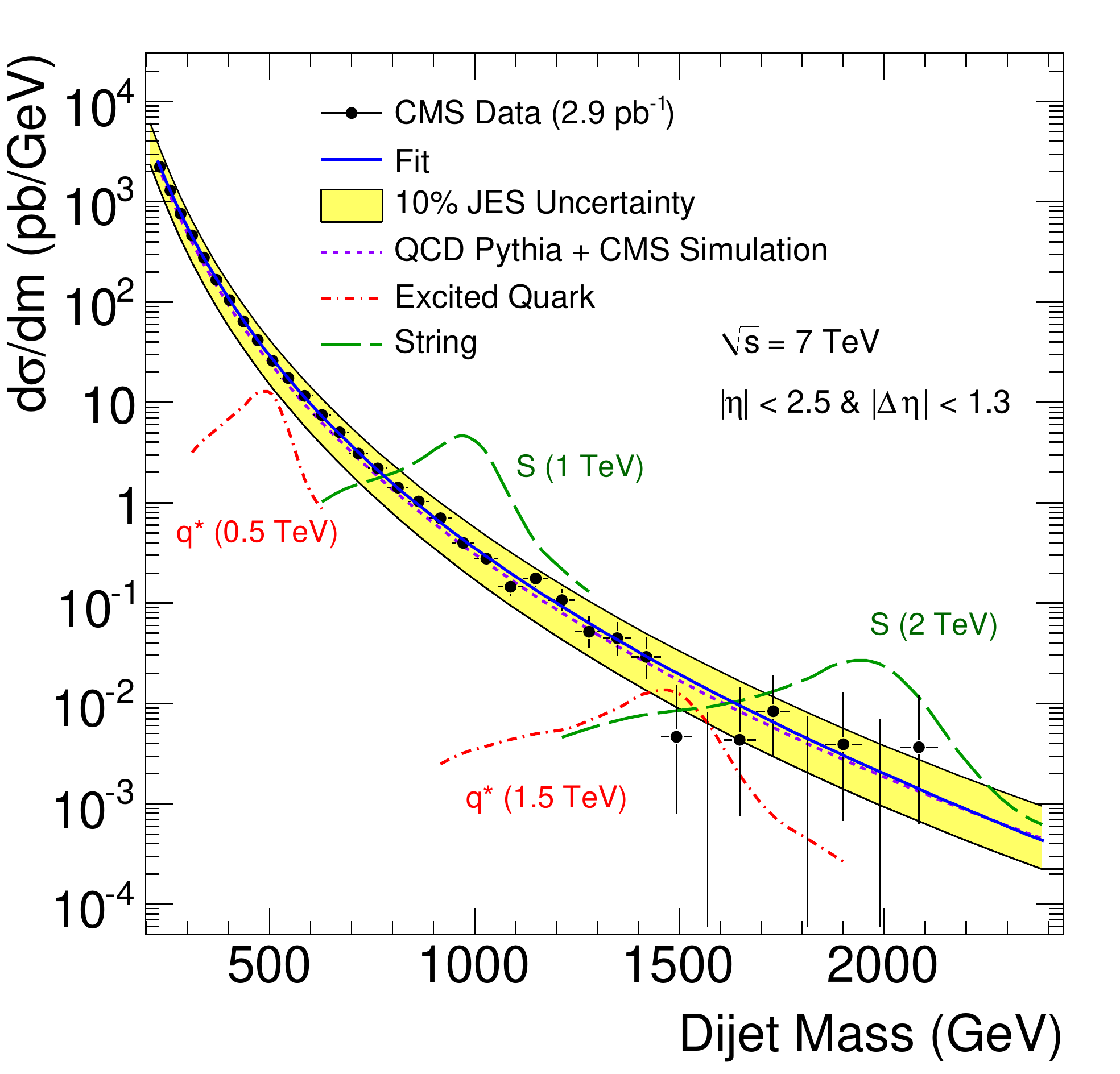}
    \hspace{1cm}
    \caption{
   Dijet mass spectrum (points) compared to a smooth fit (solid)
    and to predictions~\cite{refPYTHIA} including detector simulation of QCD (short-dashed), excited quark signals (dot-dashed),
    and string resonance signals (long-dashed). The errors are statistical only.
     The shaded band shows the effect of a 10\% systematic uncertainty in
     the jet energy scale (JES).}
    \label{fig1}
  \end{center}
\end{figure}

\begin{figure}[hbtp]
  \begin{center}
    \includegraphics[width=1.0\columnwidth]{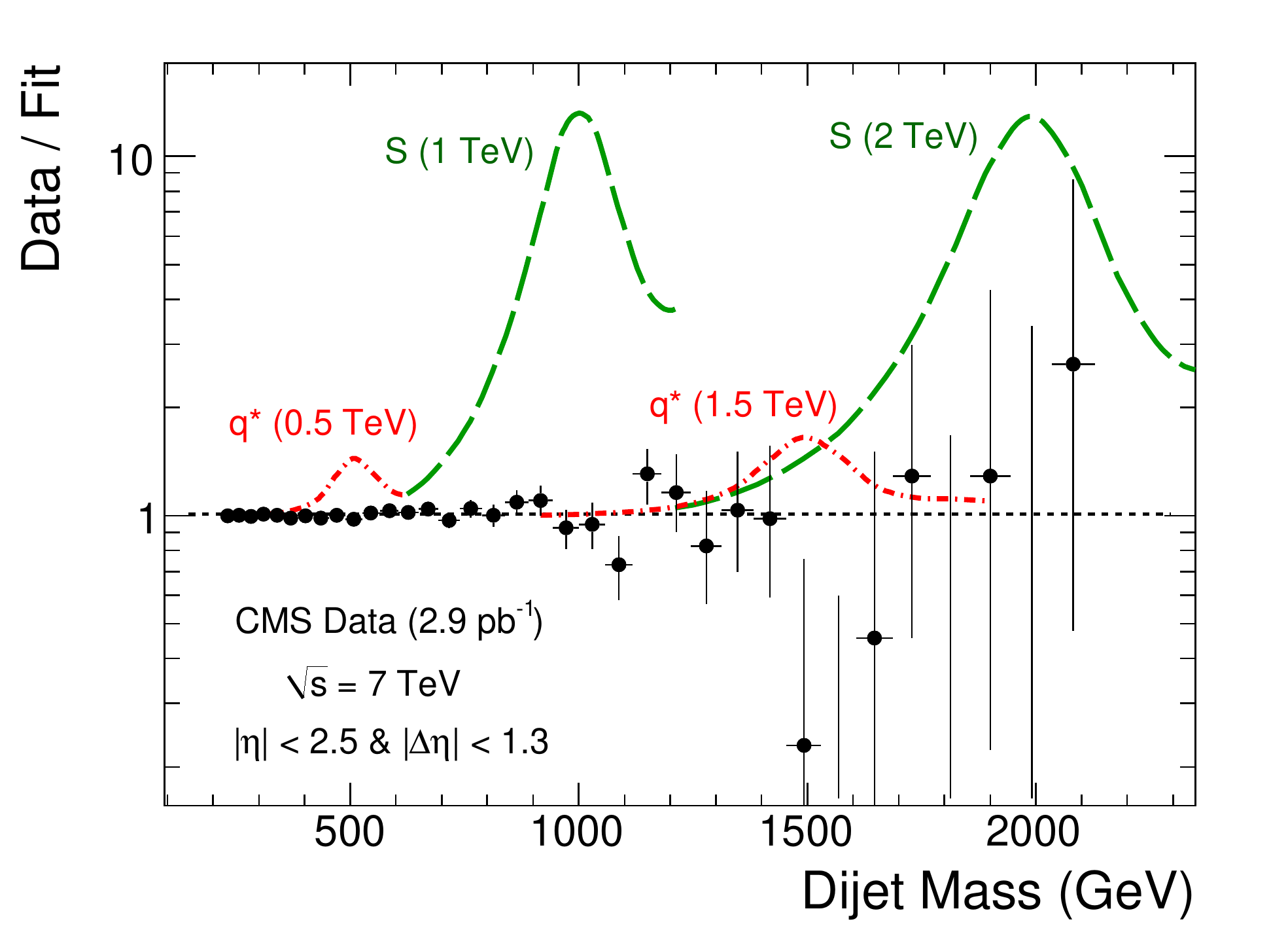}
    \hspace{1cm}
    \caption{Ratio (points) between the dijet mass data and the smooth fit,
    compared to the simulated ratios for excited quark signals
(dot-dashed) and string resonance signals (long-dashed) in the CMS detector.
The errors are statistical only.}
    \label{fig2}
  \end{center}
\end{figure}

Figure~\ref{fig1} presents the inclusive dijet mass
distribution for $pp\rightarrow$ 2 leading jets + $X$, where $X$ can be
anything, including additional jets.
We plot the measured differential cross section versus dijet mass in
bins approximately equal to the dijet mass resolution.
The data are compared to a QCD prediction from PYTHIA~\cite{refPYTHIA}, which
includes a full GEANT simulation~\cite{refGEANT} of the CMS
detector and the jet energy corrections.
The prediction uses  a renormalization scale $\mu=\pt$ and CTEQ6L1 parton distribution functions~\cite{refCTEQ}.
The PYTHIA prediction agrees with the data within the jet energy scale uncertainty, which is the dominant
systematic uncertainty.
To test the smoothness of our measured cross section as a function of dijet mass, we
fit the data with the parameterization

\begin{equation}
\frac{{\rm d}\sigma}{{\rm d}m} =
\frac{P_{0} (1 - m/\sqrt{s})^{P_{1}}}{(m/\sqrt{s})^{P_{2} + P_{3} \ln{(m/\sqrt{s})}}}\ ,
\end{equation}

with four free parameters $P_0$, $P_1$, $P_2$ and $P_3$. This functional form
has been used by prior searches to describe both data and QCD predictions~\cite{refCDFrun2,ATLAS_search}.
In Fig.~\ref{fig1} we show both the data and the fit,
which has a $\chi^2=32$ for 31 degrees of freedom.  In Fig.~\ref{fig2} we show
the ratio between the data and the fit.  The data are well described by
the smooth parameterization.

\begin{figure}[hbtp]
  \begin{center}
    \includegraphics[width=1.0\columnwidth]{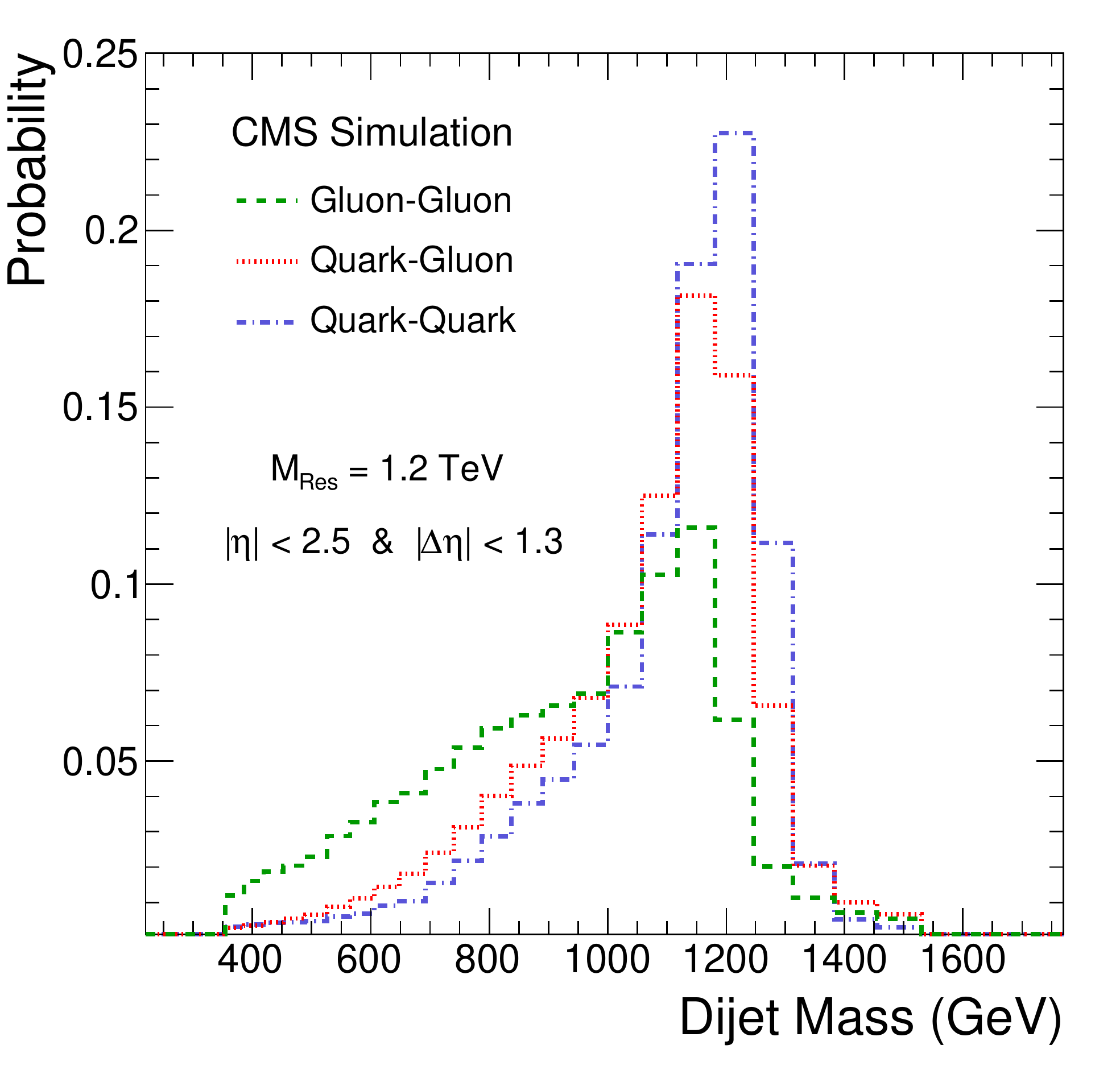}
    \hspace{1cm}
    \caption{ Simulation of the expected dijet mass distributions in the
CMS detector from a narrow 1.2~TeV
resonance of type quark-quark (dot-dashed), quark-gluon (dotted),
and gluon-gluon (dashed).}
    \label{fig3}
  \end{center}
\end{figure}

 We search for narrow resonances, for which the
natural resonance width is negligible compared to the CMS dijet mass
resolution. Figures~\ref{fig1} and \ref{fig2} present the
predicted dijet mass distribution for string resonances and excited quarks using the
PYTHIA Monte Carlo and the CMS detector simulation.
The predicted mass distributions exhibit a Gaussian core from
jet energy resolution and a tail toward low masses from QCD radiation.
This can be seen in Fig.~\ref{fig3}, which shows examples of the predicted dijet mass distribution of resonances from
three different parton pairings: $q\bar{q}$ (or $qq$) resonances
from the process $G \rightarrow q\bar{q}$~\cite{ref_rsg},
$qg$ resonances from $q^* \rightarrow qg$~\cite{ref_qstar},
and $gg$ resonances from $G \rightarrow gg$~\cite{ref_rsg}.
For resonance masses between 0.5 and 2.5 TeV, the dijet mass resolution
varies from 8\% to 5\% for $qq$, 10\% to 6\% for $qg$, and 16\% to 10\%
for $gg$, respectively.
The increase of the
width of the measured mass shape and the shift of the mass
distribution toward lower masses are enhanced when the number of
gluons in the final state is larger, because QCD radiation is larger
for gluons than for quarks.  The latter also implies that the
detector response is lower to gluon jets than to quark jets~\cite{JME-07-002-PAS} (jet
energy corrections, applied both to data and to simulations, are for
the mixture of quark and gluon jets expected in QCD). The distributions
in Fig. 3 are generically valid for other resonances with the
same parton content and with a natural width small compared to the dijet mass
resolution.
 There is no indication
of narrow resonances in our data as shown in Figs.~\ref{fig1} and ~\ref{fig2}.

We use the dijet mass data points, the background (QCD) parameterization, and the
dijet resonance shapes to set specific limits on new particles decaying to
the parton pairs $qq$ (or $q\bar{q}$), $qg$, and $gg$.
For setting upper limits,
before accounting for systematic uncertainties, we use a Bayesian
formalism with a uniform prior for the signal cross section.
We calculate the posterior probability density as a function of resonance cross section,
independently at 22 different values of the resonance mass from 0.5 to 2.6 TeV
in steps of $0.1$~TeV. From this we find initial 95\% confidence level (CL) upper limits on the cross section, including only
statistical uncertainties. The dominant sources of systematic uncertainty are the jet energy scale (10\%), the jet energy
resolution (10\%), the integrated luminosity (11\%), and the background parameterization choice (included by using a different
parameterization~\cite{Abe:1997hm} that also describes the data).
The jet energy scale and resolution uncertainties are conservative estimates,
consistent with those measured using collision data~\cite{JME-10-003-PAS}.
To incorporate systematic uncertainties, we then use an approximate technique, which in our
application is generally more conservative than a fully Bayesian
treatment.  The posterior probability density
for the cross section is broadened by convoluting it, for each resonance mass, with a Gaussian
systematic uncertainty~\cite{Abe:1997hm}.
As a result, the cross section limits including systematic uncertainties
increase by 17\%--49\% depending on the resonance mass and type.
Table~\ref{tab_limit} lists the generic upper limits at the 95\% CL on
$\sigma\times\mbox{BR}\times\mbox{A}$, the product of
cross section ($\sigma$), branching fraction (BR), and acceptance (A) for the
kinematic requirements $|\Delta\eta|<1.3$ and $|\eta|<2.5$,
for $qq$, $qg$, and $gg$ resonances. The acceptance for isotropic decays is
$\mbox{A}\approx 0.6$ independent
of resonance mass.

\begin{table}[htb]
\begin{center}
\caption{
Upper limits at the 95\% CL on $\sigma\times\mbox{BR}\times\mbox{A}$,
as a function of the new particle mass, for narrow resonances decaying to dijets
with partons of type quark-quark ($qq$), quark-gluon ($qg$), and gluon-gluon ($gg$).
The limits apply to the kinematic range where both jets have pseudorapidity
$|\eta|<2.5$ and $|\Delta\eta|<1.3$.
}
\label{tab_limit}
\vspace*{1cm}
\begin{tabular}{|c|c|c|c||c|c|c|c|}\hline
Mass   &  \multicolumn{3}{c||}{Upper Limit (pb)} & Mass   &  \multicolumn{3}{c|}{Upper Limit (pb)}\\
 (TeV) &  qq & qg & gg &  (TeV) &  qq & qg & gg\\ \hline
0.5 &  118  &  134  &  206  & 1.6 &  3.05 &  3.72 &  6.71 \\
0.6 &  182  &  229  &  339  & 1.7 &  3.13 &  3.64 &  5.88 \\
0.7 &  90.7 &  134  &  281  & 1.8 &  2.92 &  3.41 &  5.37 \\
0.8 &  70.8 &  93.5 &  177  & 1.9 &  2.73 &  3.15 &  4.78 \\
0.9 &  52.7 &  71.6 &  142  & 2.0 &  2.71 &  3.02 &  4.39 \\
1.0 &  20.3 &  29.0 &  71.4 & 2.1 &  2.50 &  2.84 &  4.15 \\
1.1 &  17.0 &  20.1 &  35.1 & 2.2 &  2.20 &  2.55 &  3.69 \\
1.2 &  17.0 &  20.4 &  32.5 & 2.3 &  1.96 &  2.28 &  3.32 \\
1.3 &  10.5 &  12.9 &  22.8 & 2.4 &  1.79 &  2.08 &  2.94 \\
1.4 &  6.77 &  8.71 &  16.4 & 2.5 &  1.67 &  1.93 &  2.74 \\
1.5 &  3.71 &  5.02 &  10.3 & 2.6 &  1.55 &  1.80 &  2.50 \\
\hline
\end{tabular}
\end{center}
\end{table}

\begin{figure}[hbtp]
  \begin{center}
    \includegraphics[width=1.0\columnwidth]{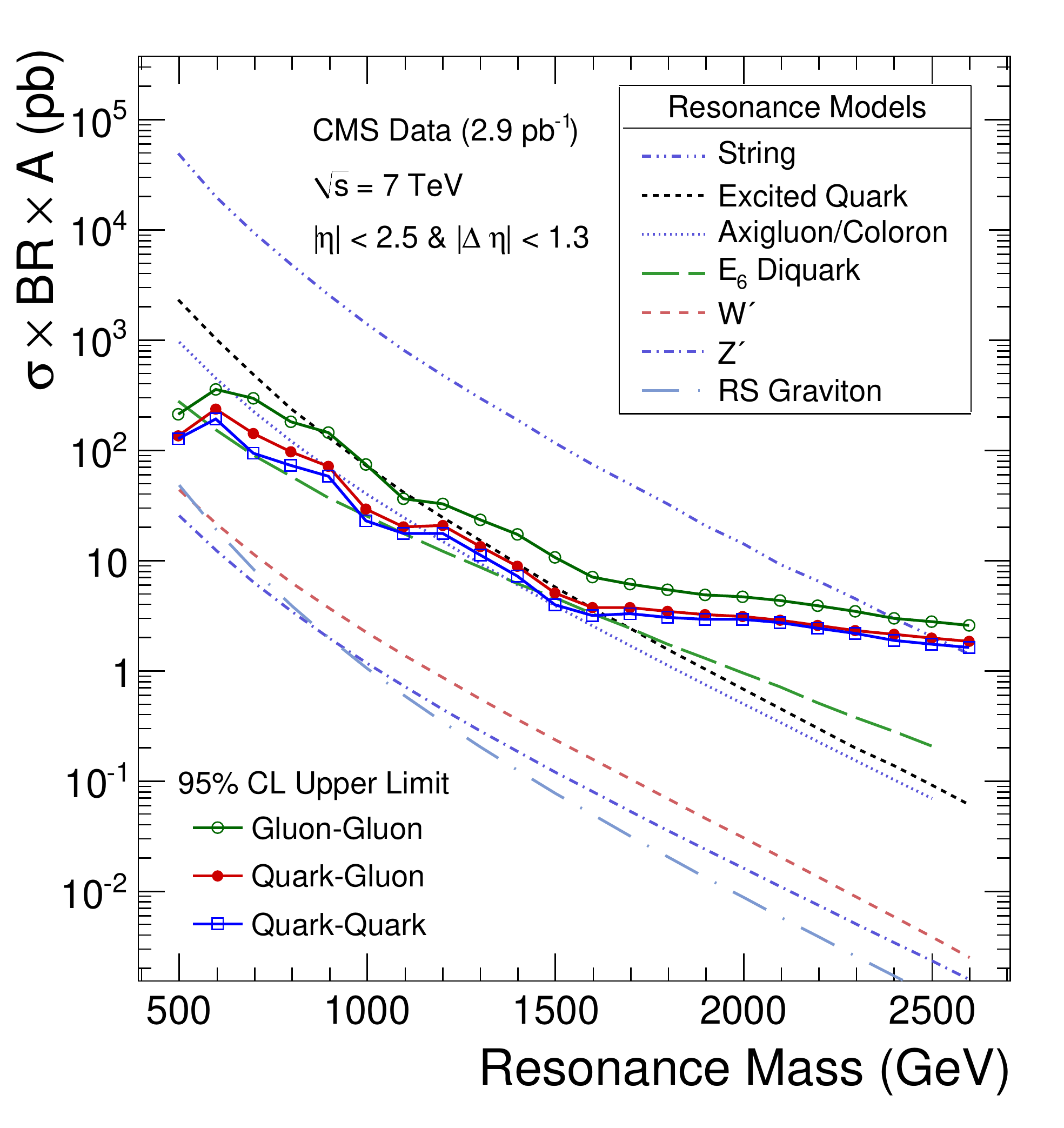}
    \hspace{1cm}
    \caption{95\% CL upper limits on $\sigma\times\mbox{BR}\times\mbox{A}$
for dijet resonances of type gluon-gluon
(open circles), quark-gluon (solid circles), and quark-quark (open boxes),
compared to
theoretical predictions for string resonances~\cite{Anchordoqui:2008di}, excited
quarks~\cite{ref_qstar}, axigluons~\cite{ref_axi},
colorons~\cite{ref_coloron}, $\mbox{E}_6$ diquarks~\cite{ref_diquark},
new gauge bosons $W^{\prime}$ and $Z^{\prime}$~\cite{ref_gauge},
and Randall-Sundrum gravitons~\cite{ref_rsg}.}
    \label{fig4}
  \end{center}
\end{figure}

In Fig.~\ref{fig4} we compare these upper limits to the model predictions as a function of resonance mass.
The predictions are from lowest order calculations of the product $\sigma\times\mbox{BR}\times\mbox{A}$
using the CTEQ6L1 parton distributions~\cite{refCTEQ}. New particles are excluded at the 95\% CL in
mass regions for
which the theory curve lies above our upper limit for the appropriate pair of
partons.
We also determine the expected lower limit on the mass of each new particle, for a smooth background in the
absence of signal.
For string resonances the expected mass limit is $2.40$~TeV, and we use the limits on $qg$ resonances to
exclude the mass range $0.50 < M(S) < 2.50$~TeV.
For comparison, previous measurements~\cite{refCDFrun2}
imply a limit on string resonances of about 1.4~TeV.
For excited quarks the expected mass limit is $1.32$~TeV, and we exclude the mass range $0.50<M(q^*)<1.58$~TeV,
extending the previous exclusion of
$M(q^*)<1.26$~TeV~\cite{Alitti:1993pn,Abe:1995jz,Abe:1997hm,refCDFrun2,Abazov:2003tj,ATLAS_search}.
For axigluons or colorons the expected mass limit is $1.23$~TeV,
and we use the limits on $qq$ resonances to exclude the mass intervals
$0.50<M(A)<1.17$~TeV and $1.47<M(A)<1.52$~TeV, extending the
previous exclusion of
$0.11<M(A)<1.25$~TeV~\cite{Albajar:1988rs,Abe:1989gz,Abe:1993it,Abe:1995jz,Abe:1997hm,refCDFrun2}. For $\mbox{E}_6$ diquarks the expected
mass limit is $1.05$~TeV, and we exclude the mass
intervals $0.50<M(D)<0.58$~TeV, and $0.97 < M(D) < 1.08$~TeV, and $1.45 < M(D) < 1.60$~TeV,
extending the previous exclusion of $0.29<M(D)<0.63$~TeV~\cite{Abe:1997hm,refCDFrun2}.
For $W^\prime$, $Z^\prime$ and RS gravitons we do not expect any mass limit, and do not exclude any mass intervals
with the present data.
The systematic uncertainties included in this analysis reduce the excluded upper masses
by roughly $0.1$~TeV for each type of new particle.

In conclusion, the measured dijet mass spectrum is a smoothly falling
distribution as expected within the standard model.
We see no evidence for new particle production. Thus we present generic upper limits
on $\sigma\times\mbox{BR}\times\mbox{A}$ that can be applied to any model
of dijet resonances,
and set specific mass limits on string resonances, excited quarks, axigluons, flavor universal colorons,
and $\mbox{E}_6$ diquarks, all of which extend previous exclusions.

\bigskip

We wish to congratulate our colleagues in the CERN accelerator departments for the excellent performance of the LHC machine. We thank the technical and administrative staff at CERN and other CMS institutes, and acknowledge support from: FMSR (Austria); FNRS and FWO (Belgium); CNPq, CAPES, FAPERJ, and FAPESP (Brazil); MES (Bulgaria); CERN; CAS, MoST, and NSFC (China); COLCIENCIAS (Colombia); MSES (Croatia); RPF (Cyprus); Academy of Sciences and NICPB (Estonia); Academy of Finland, ME, and HIP (Finland); CEA and CNRS/IN2P3 (France); BMBF, DFG, and HGF (Germany); GSRT (Greece); OTKA and NKTH (Hungary); DAE and DST (India); IPM (Iran); SFI (Ireland); INFN (Italy); NRF and WCU (Korea); LAS (Lithuania); CINVESTAV, CONACYT, SEP, and UASLP-FAI (Mexico); PAEC (Pakistan); SCSR (Poland); FCT (Portugal); JINR (Armenia, Belarus, Georgia, Ukraine, Uzbekistan); MST and MAE (Russia); MSTD (Serbia); MICINN and CPAN (Spain); Swiss Funding Agencies (Switzerland); NSC (Taipei); TUBITAK and TAEK (Turkey); STFC (United Kingdom); DOE and NSF (USA).

\bibliography{auto_generated}   

\cleardoublepage\appendix\section{The CMS Collaboration \label{app:collab}}\begin{sloppypar}\hyphenpenalty=5000\widowpenalty=500\clubpenalty=5000\textbf{Yerevan Physics Institute,  Yerevan,  Armenia}\\*[0pt]
V.~Khachatryan, A.M.~Sirunyan, A.~Tumasyan
\vskip\cmsinstskip
\textbf{Institut f\"{u}r Hochenergiephysik der OeAW,  Wien,  Austria}\\*[0pt]
W.~Adam, T.~Bergauer, M.~Dragicevic, J.~Er\"{o}, C.~Fabjan, M.~Friedl, R.~Fr\"{u}hwirth, V.M.~Ghete, J.~Hammer\cmsAuthorMark{1}, S.~H\"{a}nsel, C.~Hartl, M.~Hoch, N.~H\"{o}rmann, J.~Hrubec, M.~Jeitler, G.~Kasieczka, W.~Kiesenhofer, M.~Krammer, D.~Liko, I.~Mikulec, M.~Pernicka, H.~Rohringer, R.~Sch\"{o}fbeck, J.~Strauss, A.~Taurok, F.~Teischinger, W.~Waltenberger, G.~Walzel, E.~Widl, C.-E.~Wulz
% ...
\vskip\cmsinstskip
% ...
1:~~Also at CERN, European Organization for Nuclear Research, Geneva, Switzerland\\
% ...

\end{sloppypar}
\end{document}